    \documentstyle[proceedings,psfig,namedreferences]{crckapb} 
\begin{opening}
\title{CHEMICAL YIELDS \protect\\
       FROM LOW- AND INTERMEDIATE-MASS STARS}
\author{P. MARIGO}
\institute{Department of Astronomy, University of Padova\\
           Vicolo dell'Osservatorio 5, 35122 Padova, Italy }
\end{opening}

\runningtitle{Yields from low- and intermediate-mass stars}

\begin{document}

% The \begin{document} command comes after the \end{opening}
% command.

\section*{Abstract}
We present new  sets of chemical yields from low- and intermediate-mass
stars with $0.8 M_{\odot}\le M \le M_{\rm up} \sim 5 M_{\odot}$,
and three choices of the metallicity, $Z=0.02$, $Z=0.008$,
and  $Z=0.004$ (Marigo 2000, in preparation). 
These are then compared with the yields calculated by other
authors 
on the basis of different model prescriptions, and basic observational
constraints which should be reproduced.  

\section{Surface chemical abundances}
In this work we predict the changes in the surface abundances 
of several elements 
(H, $^{3}$He, $^{4}$He, $^{12}$C, $^{13}$C, $^{14}$N, $^{15}$N, 
$^{16}$O, $^{17}$O, and $^{18}$O), for a dense grid of stellar models with
initial masses from $\sim 0.8 M_{\odot}$ to $M_{\rm up} = 5 M_{\odot}$, 
and three
choices of the original composition, i.e. 
$[Y=0.273, Z=0.019]$,
$[Y=0.250, Z=0.008]$, 
$[Y=0.240, Z=0.004]$.
 
Various processes may concur to alter the surface chemical
composition of a star, namely: i) 
the first dredge-up occurring at the base of the RGB; 
ii) the second dredge-up taking place during the E-AGB only for stars
with $M > (3.5 - 4.0) M_{\odot}$;
iii) the third dredge-up  experienced  by  stars with
$M > (1.2 - 1.5) M_{\odot}$ during the TP-AGB phase;
and iv) hot-bottom burning in the most massive AGB stars with 
$M > (3.5 - 4.0) M_{\odot}$.

Predictions for  
first and second dredge-up are taken from Padova stellar models 
with overshooting (Girardi {\it et al.} 2000), whereas for the TP-AGB
phase the results of synthetic 
calculations are adopted (Marigo {\it et al.} 1996, 1998, 1999). 
The reader should refer to these works for all the details.
\begin{figure}
%\vspace{12truecm}
\psfig{file=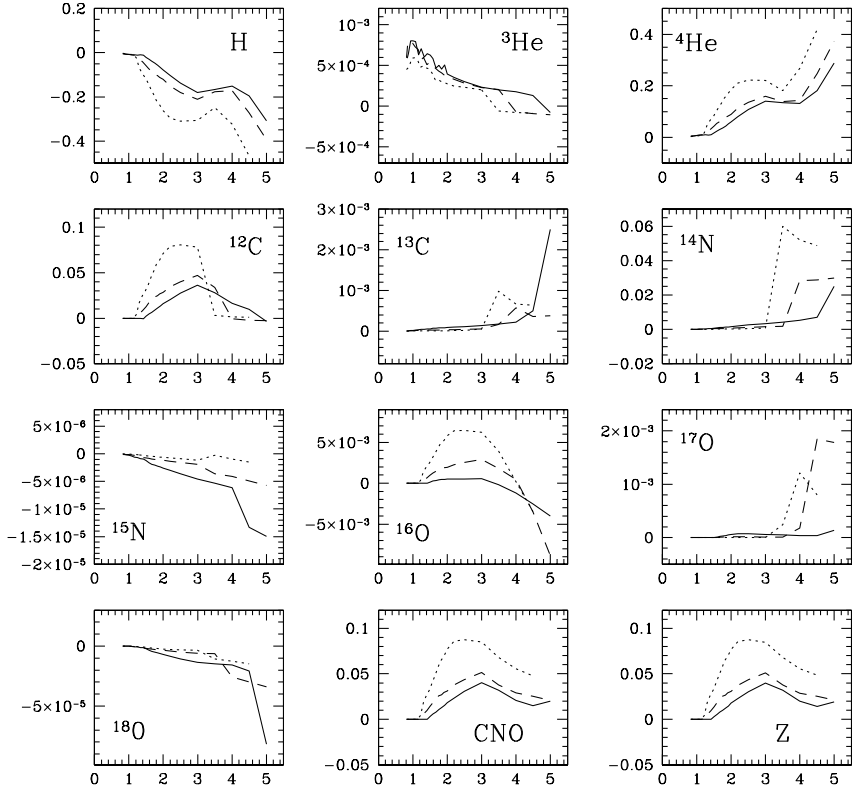,height=12truecm}
\caption{ Net chemical yields (in $M_{\odot}$)  
as a function of the initial mass (in $M_{\odot}$) of the star.
The solid, dashed, and dotted lines correspond  to metallicities
$Z=0.019$, $Z=0.008$, and $Z=0.004$, respectively.
The adopted mixing length parameter is $\alpha=1.68$.}
\label{yield}  
\end{figure}
\section{Stellar yields}
Yields from low- and intermediate-mass stars are determined 
by the wind contributions during the RGB and AGB phases. In these 
calculations mass loss 
is described by 
the Reimers' prescription ($\eta=0.45$) for the RGB phase, and by the 
Vassiliadis \& Wood (1993) formalism for the AGB phase.
Yields for the elements under consideration are shown 
in Fig.~\ref{yield} as a function of the initial stellar mass,
for three choices of the metallicity.
Major positive contributions correspond to $^{4}$He,
$^{12}$C, and $^{14}$N.
Complete sets of stellar yields, 
distinguishing both the secondary and primary
components of the CNO contributions, will be available in Marigo 
(2000, in preparation).

\section{Comparison with previous synthetic AGB models}
\begin{figure}
%\vspace{12truecm}
\psfig{file=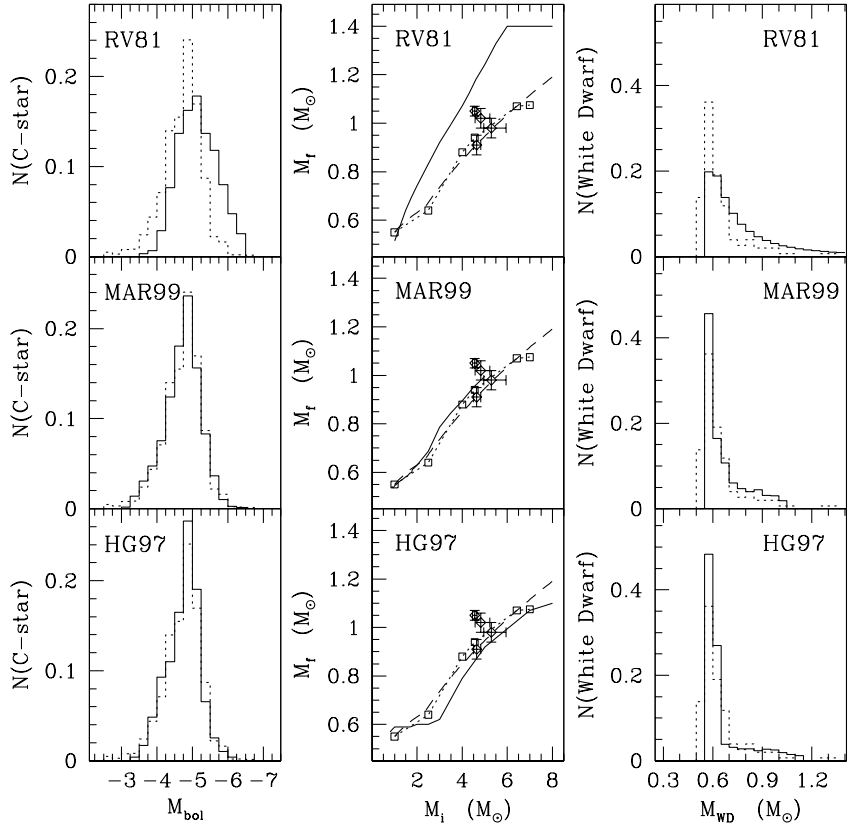,height=12.truecm}
\caption{Constraints on AGB models.
Observations are shown with dotted and dashed lines.
Theoretical predictions of different authors are presented with solid lines.
Left panels: The carbon star luminosity function in the LMC.
The observed data are from Costa \& Frogel (1996).
Middle panels: The initial-final mass relation in the solar neighbourhood.
The observed data are taken from Herwig 1996 (empty squares 
connected by dotted line), and Jeffries 1997 
(squares with error bars).  
Right panels: The white dwarf mass distribution in the solar neighbourhood.
Observations are from Bergeron {\it et al.} (1992), 
and Bragaglia {\it et al.} (1995). 
}
\label{observ}
\end{figure}
\begin{figure}[]
%\vspace{4.5truecm}
\psfig{file=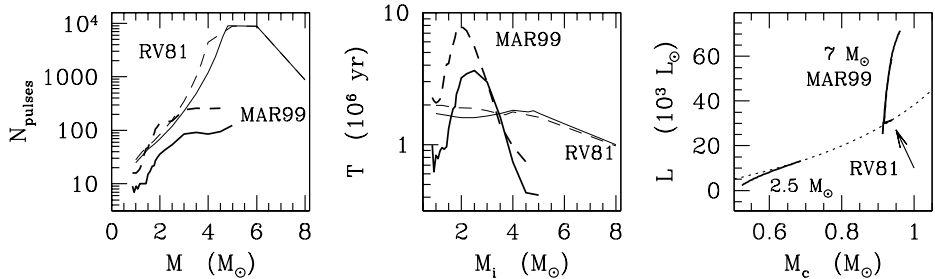,height=4.5truecm}
\caption{Comparison of model predictions. 
Left panel: Number of thermal pulses as a function of the initial
stellar mass, for metallicities $Z=0.02$ (solid line), and 
$Z=0.004$ (dashed line).
Middle panel: TP-AGB lifetimes as a function of the stellar initial mass.
Right panel: Luminosity evolution of a $7 M_{\odot}$ TP-AGB model
with hot-bottom burning as a function of the core mass (solid lines),
as predicted by MAR99 and RV81 (according to their Eq.~(15) for the 
same $M_{\rm c}$).
The standard $M_{\rm c} - L$ relation, followed by
the 2.5 $M_{\odot}$ model,  
is shown by the dotted line.}
\label{model}
\end{figure}

Chemical yields crucially depend on the adopted mass-loss and  
nucleosynthesis prescriptions.
In Fig.~\ref{observ} a comparison is made between 
different synthetic AGB models - namely: 
Renzini \& Voli 1981 (RV81), van de Hoek \& Groenewegen 1997
(HG97), and this work (MAR99) - on the basis
of some key-observables.
All these constraints are satisfactorily reproduced by
both calibrated HG97 and MAR99 models, whereas 
RV81 results are quite discrepant.
Specifically, the uncalibrated RV model 
predicts too few faint and too many bright carbon stars, 
together with  a sizeable excess of massive white dwarfs 
(with $M  > 0.7 M_{\odot}$). This can be explained
considering the lower efficiency of both the third dredge-up and mass-loss 
(Reimers' law  with $\eta=0.333 - 0.666$) adopted by RV81. 

The effect of different mass-loss prescriptions is also 
clear from Fig.~\ref{model}.
The most massive AGB stars are expected 
to suffer a huge number of thermal pulses 
(hence dredge-up episodes) in the RV81 model ($\sim 10^4$), 
about two order of magnitude more than in the MAR99 model ($\sim 10^2$,
left panel).
As a consequence, the duration of the TP-AGB phase for these models 
is affected in the same direction, being
much longer in RV81 than in MAR99 (middle panel 
of Fig.~\ref{model})

A final remark concerns the process of hot-bottom burning suffered
by the most massive AGB stars ($M > (3.5 - 4)\, M_{\odot}$). 
Its treatment determines the temperature at the base 
of the convective envelope and the related 
nucleosynthesis, as well as the luminosity evolution of these stars.
To this respect, it is worth noticing 
that the break-down of the $M_{\rm c}-L$ relation 
(first pointed out by Bl\"ocker \& Sch\"onberner 1991) is included   
by MAR99 (see Marigo 1998), 
whereas in RV81 and HG97 this overluminosity effect 
is not taken into account (right panel of Fig.~\ref{model}).

\begin{figure}
\begin{minipage}{0.63\textwidth}
\psfig{file=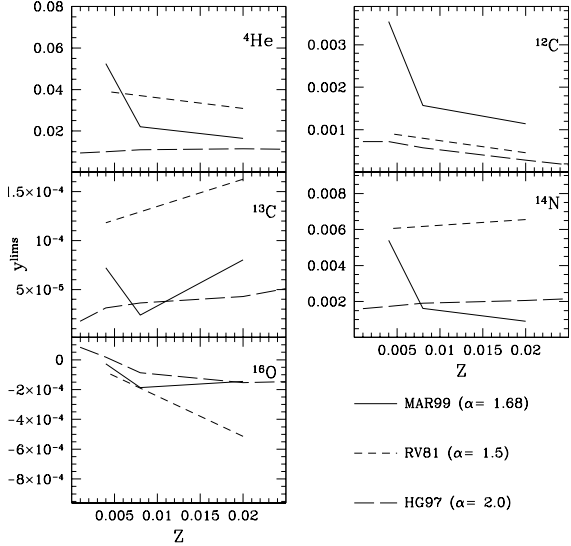,width=\textwidth}
\end{minipage}
\hfill
\begin{minipage}{0.36\textwidth}
\caption{\mbox{Integrated yield} contributions from low- and intermediate-mass
stars as a function of the metallicity, as defined by Eq.\protect(\ref{sspy}).
The mixing-length parameters ($\alpha$) adopted by the authors are indicated.}
\label{sspy}
\end{minipage}
\end{figure}

\section{Yields from single stellar populations}
In order to compare stellar yields with different values of
$M_{\rm up}$ (i.e. $8 M_{\odot}$ for RV81 and HG97
classical models, and $5 M_{\odot}$ for MAR99 model with overshooting),
we calculate the quantities   
\begin{equation}
y^{\rm lims}_{k} = \frac{ \int_{0.8}^{M_{\rm up}} m p_{k}(m) \phi(m) dm }
		   { \int_{0.8}^{100} m \phi(m) dm }
\label{ylims}
\end{equation}
where $p_{k}(m)$ is the fractional yield of the element
$k$ produced by a star of initial mass $m$. 
These quantities express the relative  chemical contribution  
from low- and intermediate stars belonging
to a given simple stellar population.
They are shown in Fig.~\ref{sspy} as a function of the metallicity for
the three sets here considered.

Differences show up both in metallicity trends and 
absolute values of $y^{\rm lims}_{k}$.
Compared to previous calculations, MAR99 yields show
a pronounced dependence on the metallicity, i.e. yields increase with 
decreasing $Z$. Conversely, the RV81 and HG97 sets 
present quite weak trends with $Z$.

The metallicity dependence can be explained as follows.
On one side, AGB lifetimes of low-mass stars increase
at decreasing  metallicities, as mass-loss rates are expected to be lower.
This fact leads to  a larger number of dredge-up episodes.
Moreover, both the onset and the efficiency of the third dredge-up
are favoured at lower metallicities. These 
factors concur to produce a greater enrichment in carbon.  
On the other side, hot-bottom burning in more
massive AGB stars becomes more efficient
at lower metallicities, leading to a greater enrichment in nitrogen.
The combination of all factors favours higher positive yields
of helium at lower $Z$.

As far as the single elemental species are concerned, we can notice:
\begin{itemize}
\item MAR99 yields of $^4$He are larger than those by HG97, due to 
the earlier activation of the third dredge-up and, likely, 
to a greater efficiency/duration of hot-bottom burning in our models.
Predictions by RV81 show no significant trend with $Z$, 
and higher positive yields 
(due to the quite low mass-loss rates adopted).
\item MAR99  yields of $^{12}$C are systematically higher than those
of RV81 and HG97 because of the 
 earlier onset (and average greater efficiency than in RV81) 
of the third dredge-up.
\item The dominant contribution to the yields of $^{14}$N comes
from hot-bottom burning in the most massive AGB stars.
Differences in the results reflect  different efficiencies
of nuclear reactions and AGB lifetimes.
In particular, according to MAR99 the production of $^{14}$N,
mainly of primary synthesis, 
is favoured at lower $Z$. 
\end{itemize}

\end{document}